# Micro-Structured Two-Component 3D Metamaterials with Negative Thermal-Expansion Coefficient from Positive Constituents

Jingyuan Qu[1,2], Muamer Kadic[1,2], Andreas Naber[1], and Martin Wegener[1,2]

[1] *Institute of Applied Physics, Karlsruhe Institute of Technology (KIT), 76128 Karlsruhe, Germany*

[2] *Institute of Nanotechnology, Karlsruhe Institute of Technology (KIT), 76128 Karlsruhe, Germany*

**Abstract**

Controlling the thermal expansion of materials is of great technological importance. Uncontrolled thermal expansion can lead to failure or irreversible destruction of structures and devices. In ordinary crystals, thermal expansion is governed by the asymmetry of the microscopic binding potential, which cannot be adjusted easily. In artificial crystals called metamaterials, thermal expansion can be controlled by structure. Here, following previous theoretical work, we fabricate three-dimensional (3D) two-component polymer microlattices by using gray-tone laser lithography. We perform cross-correlation analysis of optical microscopy images taken at different sample temperatures. The derived displacement-vector field reveals that the thermal expansion and resulting bending of the bi-material beams leads to a rotation of the 3D chiral crosses arranged onto a 3D checkerboard pattern within one metamaterial unit cell. These rotations can over-compensate the expansion and lead to an effectively negative thermal length-expansion coefficient – for all positive constituents – evidencing a striking level of thermal-expansion control.

**Introduction**

Three-dimensional (3D) printing of materials is a huge trend. It allows for individualizing products and for fabricating architectures that are very difficult if not impossible to make otherwise. Ultimately, one would like to 3D print any functional structure or device at the push of a button. Apart from boosting spatial resolution and printing speed, achieving this goal requires the ability to obtain hundreds or thousands of different material properties with one 3D printer. Today's 2D graphical printers realize thousands of colors from only three cartridges (cyan, magenta, yellow). By analogy, future 3D material printers might be able to print thousands of different effective materials from only a few constituent-material "cartridges".

Physics is on our side: Upon 3D printing two constituent materials **A** and **B** to obtain a composite or metamaterial, one might naively think that its effective properties will always be in between those of **A** and **B**. Fortunately, this is *not* the case.[1-4] In some cases, the behavior is even conceptually unbounded, i.e., an effective material parameter can assume any value from minus infinity to plus infinity even if those of the constituents are all finite and, e.g., positive. Examples are the electric permittivity and the magnetic permeability in electromagnetism or the compressibility and the mass density in mechanics.[5-13] However, for the mentioned examples, sign reversal and unbounded effective parameters are only possible near resonances at finite frequency and not in the truly static regime for reasons of stability in mechanics and non-negative energy density in electromagnetism.[5,13]

Static examples are rare. Theoretically, the thermal length-expansion coefficient and the Hall resistance have been discussed.[1-3,14-21] Regarding the Hall resistance, even one constituent material **A** and voids within suffices.[20] The situation is distinct for the thermal length-expansion coefficient. Within the range of validity of the continuum approximation, any connected structure composed of one constituent material **A** and voids within will show exactly the same thermal length-expansion coefficient as the bulk constituent material **A**. In contrast, the work of Lakes and others has shown that the behavior of composites containing *two* components **A** and **B** plus voids within is principally unbounded.[1,3,14-16]

In regard to applications, thermal length-expansion is a small effect with huge consequences. A relative thermal length-expansion around $10^{-4}$ to $10^{-3}$ can lead to severe misalignment, failure, or cracks. Atomic-scale composites can provide near-zero or negative thermal-length expansion by changing the microscopic binding potential.[22-24] More macroscopic composites with near-zero length expansion are based on one constituent material with positive and another one with negative thermal expansion. For example, CERAN® glass cooking fields are made like that and have led to considerable markets.

**Results**

In this work, by using 3D gray-tone two-photon laser lithography, we fabricate micro-structured two-component metamaterials using a single photoresist, leading to an effectively negative thermal length-expansion coefficient from all-positive constituents. Applying image cross-correlation analysis, we directly measure the temperature-induced displacement-vector field in different layers of the micro-lattice with sub-pixel precision and thereby visualize the underlying microscopic mechanism.

Figure 1a exhibits a single lattice constant of the micro-lattice blueprint we start from. This unit cell is placed onto a three-dimensional simple-cubic translational lattice. Apart from minor modifications, this blueprint has been taken from the literature.[16] The two components **A** and **B** shown in different colors have different positive thermal length-expansion coefficients. Intuitively, the operation principle is as follows (see Fig. 1b): The bi-material beams expand and bend upon heating. The bending leads to a rotation of the 3D crosses, the arms of which make them chiral. The chirality and hence the sense of rotation alternates between clockwise and counter-clockwise from one 3D cross to its neighbors, forming a 3D checkerboard pattern. The rotations counteract the length expansion of the beams. Thus, for sufficiently pronounced rotations, the crossing points move towards each other and the effective thermal length-expansion of the micro-lattice becomes negative, i.e., the lattice contracts. For somewhat less pronounced rotations, zero thermal length-expansion results (see Supplementary Fig. 1). If component **B** is left away, i.e., replaced by vacuum, the effective thermal length-expansion coefficient of the micro-lattice is identical to that of bulk **A**. The same is true if **B** equals **A**.

We mention in passing that the structure shown in Fig. 1a has an effective Poisson's ratio of $v = -0.41$. However, in general, the sign of the thermal-length expansion coefficient is not necessarily the same as the sign of the Poisson's ratio.[15]

How can we fabricate such complex 3D two-component metamaterials? Three-dimensional two-photon laser printing of a single polymer component is a well-established technology. Here, we use a commercial instrument (Photonics Professional, Nanoscribe GmbH). While characterizing the thermal length-expansion coefficients of various polymers in bulk cuboid form, we noticed that the thermal length-expansion coefficient does not only depend on the type of monomer we start from. For a given monomer (we use IP-Dip, Nanoscribe GmbH), it also depends on the light exposure dose during photo-polymerization. For example, when writing a bulk cube at a power scaling factor (see Experimental) of 60% (40%) at a fixed scan speed of 2 cm/s, we find a thermal length-expansion coefficient of $\alpha_L = +5 \times 10^{-5}\ \text{K}^{-1}$ ($\alpha_L = +9 \times 10^{-5}\ \text{K}^{-1}$). The measurement procedure shall be described below. We interpret this behavior in that the exposure dose influences the polymer cross-linking density. A correlation between cross-linking density and thermal expansion has been established previously.[26] The bottom line of this finding is that we can realize two different components **A** and **B** plus voids within by using only a single photoresist and gray-tone optical lithography.

For the temperature-dependent measurements, the samples are fixed to a Peltier-element heater within a small encapsulated chamber. In this fashion, the air within the chamber is heated as well, such that, after a certain waiting time (we chose several hours here), we can safely assume that the polymer micro-lattice temperature is actually equal to the temperature measured by a calibrated thermo-resistor at the Peltier-element location (as well as by a second one at the upper end of the cell). The chamber has a glass window to allow for optical access via a home-built wide-field optical microscope. It is based on a single microscope lens (Zeiss LD Achroplan 20 × with a numerical aperture $NA = 0.4$) which images a sample plane directly onto the chip of a silicon-based charge-coupled-device (CCD) camera. We illuminate the sample by diffuse white light, for which we obtain the best image contrast for our micro-lattices. The entire chamber can be moved with respect to the microscope by using a 3D piezoelectric translation stage (Piezosystem Jena 1469), which is controlled by a computer.

How can we directly characterize the thermal length-expansion and the operation principle underlying the micro-lattices? For example, for temperature differences on the order of $\Delta T = 20$ K, we expect relative length changes of the samples on the order of $|\Delta L/L| = 10^{-3}$. If one images the entire sample onto a camera chip with 1000 pixels in one direction, this relative change corresponds to a movement of merely 1 pixel. Obviously, the measurement accuracy must be yet better than that. Image cross-correlation analysis can provide such sub-pixel sensitivity.[25] Our analysis starts from two optical images of a sample taken at two different temperatures. The images can refer to the sample's surface. For transparent samples, they can alternatively correspond to a plane below the surface within the volume of the sample (e.g., plane $P_1$ or $P_2$ in Fig. 1c). In one image taken at room temperature, regions of interest (ROI) are defined. These small regions ($33 \times 33$ pixels) can be positioned onto characteristic points of the sample, e.g., onto the lattice points of a micro-lattice (compare Fig. 1c). A second image is taken at an elevated temperature. Next, the two-dimensional cross correlation of the two images is computed. If both images are identical, the cross correlation peaks at displacement vector (0,0); if they are shifted with respect to each other, it peaks at the corresponding displacement vector $\vec{u} = (u_x, u_y) \neq (0,0)$. Thus, by identifying the maximum of the cross-correlation function in two dimensions, the local displacement vector can be identified. We average over the results obtained from ten different images at elevated temperature. By repeating the procedure for many ROI at coordinates $(x, y)$ in the image, we obtain the displacement-vector field $\vec{u}(x, y)$. Generally, upon changing the temperature of the sample and its holder, the sample as a whole can also move or drift across the image and/or through the focal plane of the microscope. The former effect can be eliminated by subtracting the average displacement vector of all ROI, i.e., $\vec{u}(x, y) \rightarrow \vec{u}(x, y) - \langle \vec{u} \rangle$. This subtraction is unproblematic if the camera chip is perfectly homogeneous. To avoid possible artifacts from inhomogeneities, we rather compensate most of this average displacement by the piezoelectric translation stage, such that the sample's image on the camera chip does not change with respect to the camera chip. Remaining small average shifts are subtracted in the processing of the images. Likewise, we compensate a possible thermally induced defocus (i.e., movement in $z$-direction) by an autofocus function. It is based on acquiring images while moving the sample along the $z$-axis using the piezoelectric translation stage. The sharpest image is selected in the post-processing.

Electron micrographs of selected fabricated samples on glass substrates are shown in Fig. 2. The structure in Fig. 2b is composed of two different components, corresponding to different exposure doses during the printing process (see above). Note that the beams are bent, whereas they are straight for the control sample (depicted in Fig. 2a), in which exclusively the first component has been written. The bending originates from the different cross-linking densities resulting from different exposure doses, leading to different levels of volume shrinkage of the constituents during development. Notably, this shrinkage leads to an overall expansion of the metamaterial sample – in agreement with the operation principle.[27] This pre-bending also directly evidences that the two components actually have different properties. Within the linear regime, the pre-bending does not change the operation principle compared to that of the blueprint (compare Fig. 1b). The samples shown in Fig. 2 contain $4 \times 4 \times 2 = 32$ complex three-dimensional unit cells following the blueprint shown in Fig. 1a.

Figure 3 summarizes measurements of the displacement-vector field induced by a temperature rise of $\Delta T = 20 \text{ K}$ with respect to room temperature. The background shows the room-temperature optical image as recorded by the CCD camera, corresponding to plane $P_1$ in Fig. 1c. The yellow arrows are the displacement vectors at the 3D crossing points. To make them visible, the length of the arrows has been multiplied by a common factor with respect to the optical image as indicated. Obviously, all yellow arrows in Fig. 3c roughly point towards the center of the sample. Furthermore, the length of the arrows roughly increases linearly from the center to the sides. These two observations indicate a nearly homogeneous and isotropic behavior of the 3D crossing points. It is thus meaningful to describe the metamaterial by an effective thermal length-expansion coefficient. Upon reducing the temperature difference to $\Delta T = 10$ K, we find essentially the same behavior as for $\Delta T = 20$ K, albeit with worse signal-to-noise ratio (not depicted). From these data at $\Delta T = 20\ K$ ($\Delta T = 10\ K$), we extract an average effective metamaterial thermal length-expansion coefficient of $\alpha_L = -5 \times 10^{-5}$ K$^{-1}$ ($\alpha_L = -5 \times 10^{-5}$ K$^{-1}$). When imaging the next lower plane of crossing points (compare $P_2$ in Fig. 1c), we again find essentially the same behavior (average $\alpha_L = -5 \times 10^{-5}$ K$^{-1}$). In contrast, the single-component control sample depicted in Fig. 3a and b exhibits a positive value of $\alpha_L = +4 \times 10^{-5}$ K$^{-1}$. All of the described aspects are reversible, i.e., we have observed them many times on a given sample. We have also seen them on several samples. Our analysis also allows for quantifying deviations from ideal isotropy. We find that the modulus of the expansion coefficient is 25% larger along the $y$-direction as compared to the $x$-direction for the two-component sample. This anisotropy is assigned to remaining fabrication imperfections.

To further investigate whether the samples actually follow the behavior anticipated by the design as discussed above, we also track other points within the unit cell (compare plane $P_1$ in Fig. 1c). The blue arrows in Fig. 3d show that the connecting beams actually expand upon heating. Here, for clarity, their average displacement vector has been subtracted. A similar behavior is found for all other beams as well (not depicted). The red arrows reveal the anticipated rotation of the crosses, with alternating clockwise and counter-clockwise sense of rotation. Again, for clarity, the average displacement vector for each cross has been subtracted from the corresponding set of four arrows. Note that the red rotation vectors are generally longer than the blue ones. As explained above, this behavior leads to the negative thermal length-expansion: The shrinkage induced by the rotations over-compensates the expansions of the individual beams. For the control sample in Fig. 3b, no significant rotations are found within the noise (hence the red arrows are not depicted here) and the thermal expansion is positive – as for the bulk polymer constituents discussed above. The measured behavior in Fig. 3 is in good overall agreement with the calculated one shown in Fig. 4.

In summary, we have fabricated and characterized micrometer-scale two-component polymer-based metamaterials exhibiting an effectively negative thermal length-expansion coefficient from positive constituents. The necessary two components plus voids have been realized by 3D gray-tone laser lithography using only a single photoresist.

**Methods**

**Sample design**

For the metamaterial design, a simple cubic unit cell was analyzed by a finite-element approach using the commercially available software package COMSOL Multiphysics and the MUMPS solver within. For the mechanical part of the problem, the thermal expansion was introduced as a volumetric stress. The constituent materials **A** (and **B**) were modeled with a Young's modulus of $E_A = 4$ GPa (and $E_B = 3$ GPa), a Poisson's ratio of $\nu_A = 0.4$ (and $\nu_B = 0.4$), and a thermal length-expansion coefficient of $\alpha_L^A = +4 \times 10^{-5}$ K$^{-1}$ (and $\alpha_L^B = +6 \times 10^{-5}$ K$^{-1}$). For the calculations shown in Fig. 1b, we used periodic boundary conditions.[28] To assess the influence of finite sample size, we also performed calculations for samples containing $4 \times 4 \times 2$ unit cells fixed to a rigid substrate (Fig. 4). Based on the two mirror planes cutting the structure through the middle, one parallel to the $xz$ and one parallel to the $yz$-plane, we have reduced the computational domain to $2 \times 2 \times 2$ unit cells fixed to a rigid substrate. The other sides are left free to move (stress-free boundary conditions). The derived effective thermal length-expansion coefficients were different from those for the periodic boundary conditions by only a few percent. Typically, each unit cell (compare Fig. 1a) was discretized into $5 \times 10^5$ tetrahedral elements.

**Sample fabrication**

The structures composed of $4 \times 4 \times 2$ unit cells were written using a liquid photoresist (IP-Dip, Nanoscribe GmbH, Germany) and a commercial three-dimensional laser lithography system (Photonics Professional, Nanoscribe GmbH). The objective lens ($63\times$, $NA = 1.4$, Carl Zeiss) was dipped directly into the photoresist. An average power of 50 mW measured at the backfocal aperture of the objective lens with a diameter of 7.3 mm was defined as the reference power. The actual power was given by this reference power times the power scaling factor, which was varied locally in the spirit of gray-tone lithography. The component with low (high) thermal length-expansion coefficient was written with a power scaling factor of 65% (37%). The line spacing in the $x$- and $y$-direction was 200 nm, that in the $z$-direction was 500 nm. Small parts were written in the galvo-scanning mode with a scan speed of 2 cm/s and stitched by using micro-positioning stages.[28] The exposed samples were developed in mr-Dev 600 for 60 minutes, transferred to acetone, followed by supercritical-point-drying in acetone.

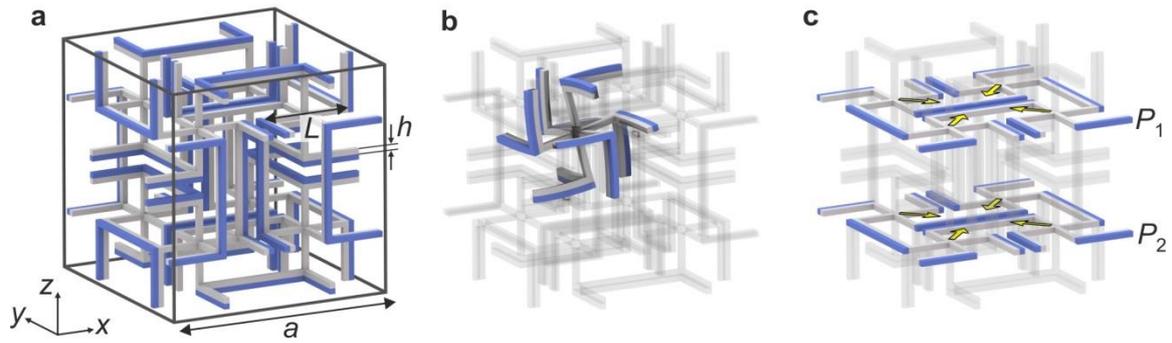

**Figure 1 | Illustration of the micro-lattice blueprint underlying our experiments. (a)** One unit cell of the three-dimensional lattice composed of two different constituent materials. The one shown in gray has a smaller positive thermal length-expansion coefficient than the one shown in blue. This unit cell is repeated on a simple-cubic translation lattice with lattice constant $a$. The distance between the 3D crosses is $a/2$. All beams have quadratic cross section with width $h$. Geometrical parameters: $a = 100$ µm, $h = 2.5$ µm, and $L = 40$ µm. **(b)** In one eighth of the unit cell, the calculated structure is exhibited for an increased temperature, assuming a linear response. For clarity, all changes are largely exaggerated. One obtains an expansion and bending of the bi-material beams. The bending leads to a rotation of the three-dimensional cross, resulting in an inward movement of this cross (negative thermal length-expansion coefficient). The other seven eighths of the unit cell behave as mirror images (with respect to the three principal Cartesian planes) of the eighth shown. **(c)** Same as panel a, but two planes $P_1$ and $P_2$ parallel to the $xy$-plane cutting through the three-dimensional crosses are highlighted. These planes are imaged by optical microscopy in the experiments shown in Fig. 3.

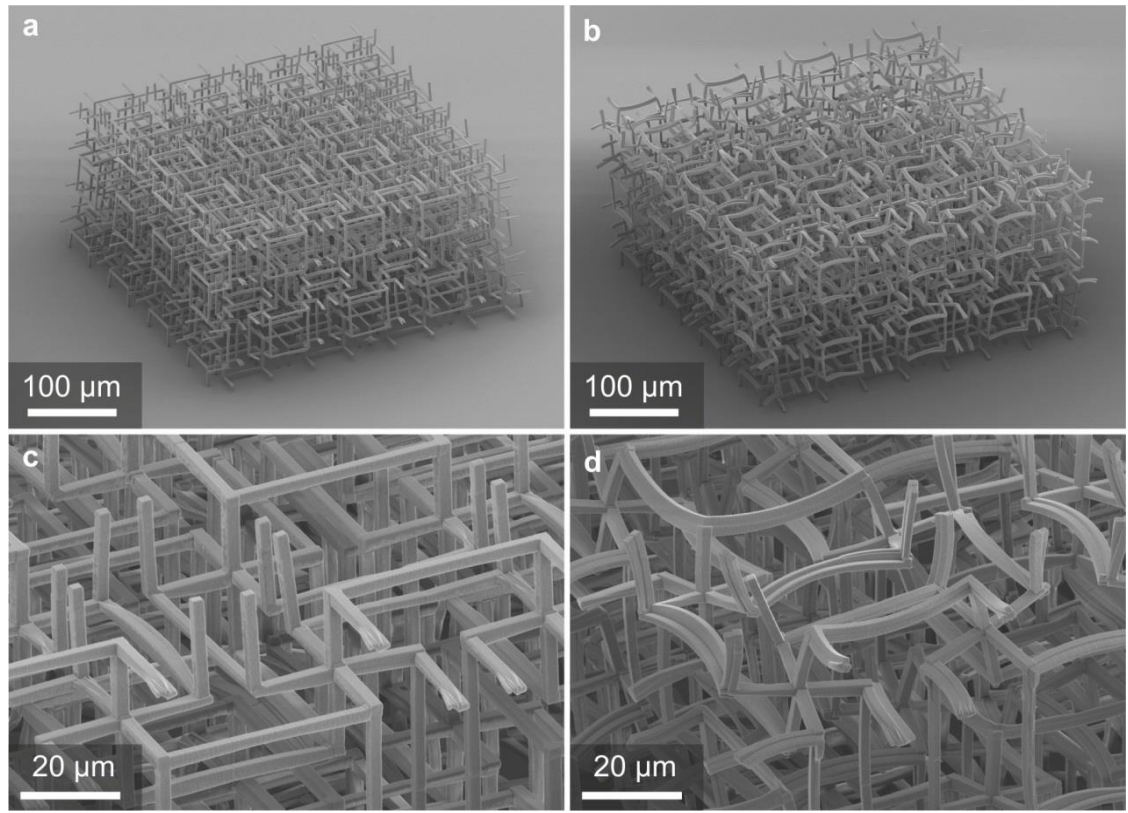

**Figure 2. | Electron micrographs of three-dimensional polymer micro-lattices fabricated by gray-tone laser lithography**. These follow the blueprint shown in Fig. 1a and contain $4 \times 4 \times 2$ unit cells. **(a)** Control sample composed of only a single constituent component (gray in Fig. 1a). **(b)** Two-component sample as shown in Fig. 1a. The different shrinkage of the two components during development leads to a pre-bending of the bi-material beams. Within the linear regime, this pre-bending does not change the operation principle. Conveniently, it directly evidences that the properties of the two components are actually different. **(c)** and **(d)** are magnified views of the samples shown in panels **a** and **b**, respectively.

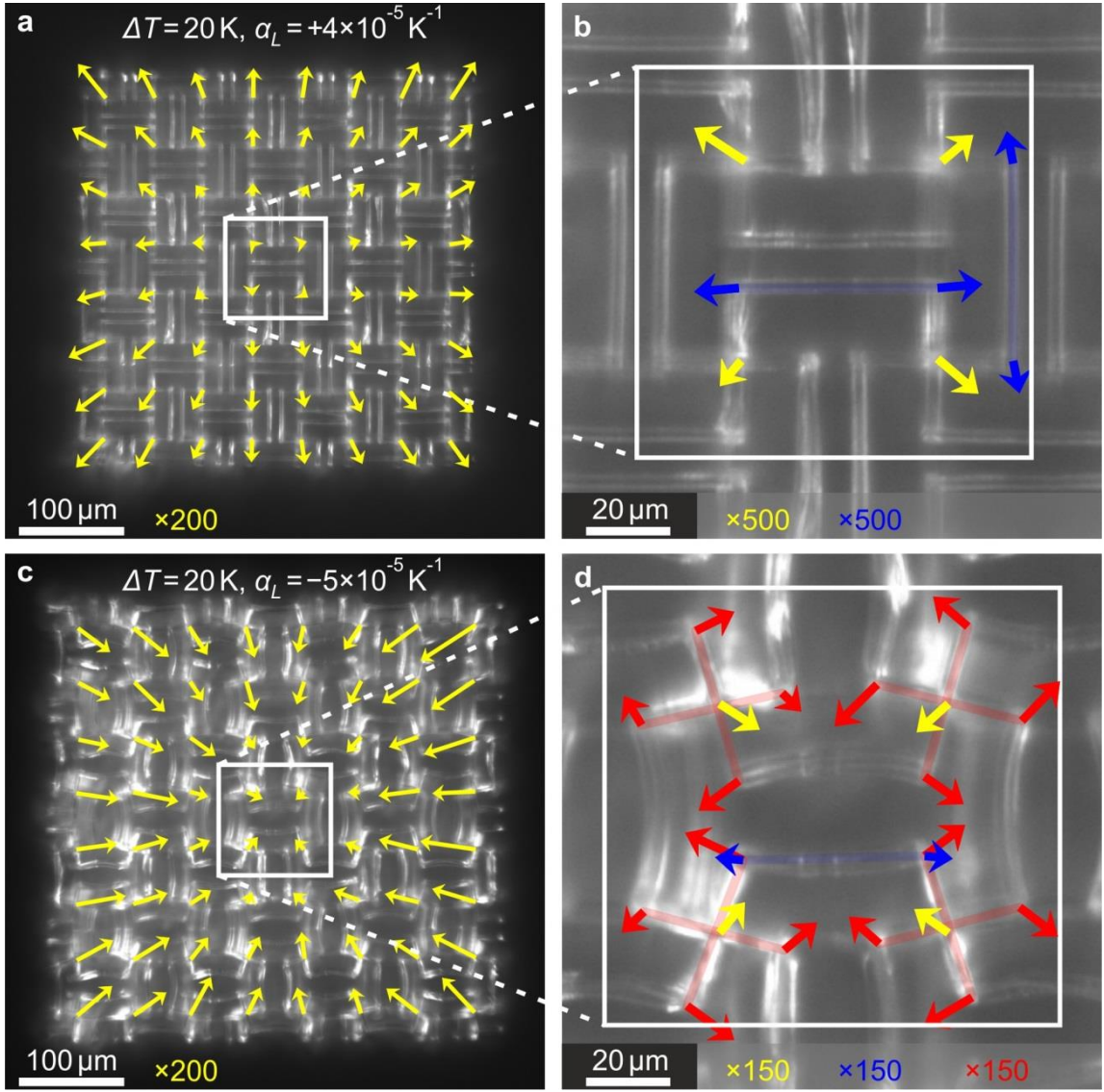

**Figure 3. | Experimental results.** Temperature-induced displacement-vector fields as measured directly by image-correlation analysis based on a room-temperature optical-microscope image (background) and images at a temperature increased by $\Delta T = 20$ K as indicated. For clarity, the lengths of the shown displacement vectors are multiplied by factors as indicated. The image plane $P_1$ is illustrated in Fig. 1c. The samples contain $4 \times 4 \times 2$ unit cells. **(a)** Control sample as shown in Fig. 2a. From the yellow displacement vectors, we derive an average thermal length-expansion coefficient of $\alpha_L = +4 \times 10^{-5}$ K$^{-1}$. **(b)** Magnified view of panel **a**. The blue vectors herein highlight the expansion of the beams. **(c)** Two-component sample (compare Fig. 1a and Fig. 2b). From the yellow displacement vectors, we derive an average thermal length-expansion coefficient of $\alpha_L = -5 \times 10^{-5}$ K$^{-1}$. **(d)** Magnified view of panel **c**. The blue vectors highlight the expansion of the

beams. Their bending leads to the rotations of the crosses as apparent from the red displacement vectors. The sense of rotation changes between clockwise and counter-clockwise in a checkerboard pattern, in agreement with the anticipated operation principle (compare Fig. 1b). The semi-transparent red and blue lines in panels **b** and **d** are guides to the eye.

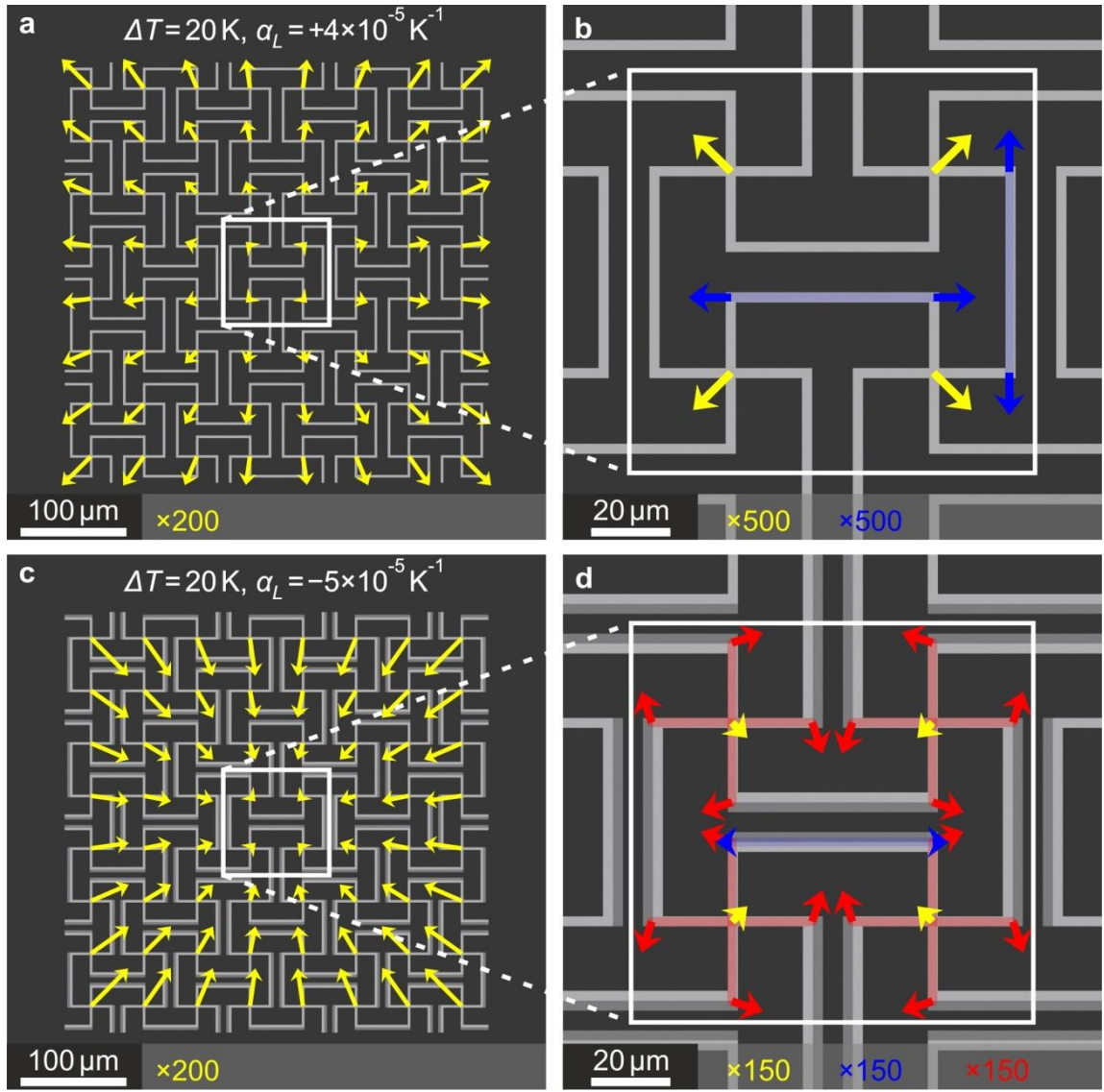

**Figure 4. | Numerical results.** Calculations corresponding to the experiments shown in Fig. 3, represented just like these experiments. Geometrical parameters (see Fig. 1a): $a = 100$ μm, $h = 2.5$ μm, and $L = 40$ μm. Material parameters: $E_A = 4$ GPa, $E_B = 3$ GPa, $\alpha_L^A = +4 \times 10^{-5}$ K$^{-1}$, and $\alpha_L^B = +6 \times 10^{-5}$ K$^{-1}$.

**Acknowledgements**

We acknowledge discussions with Claudio Findeisen and Peter Gumbsch (KIT). We are grateful to Martin Schumann (KIT) for help regarding the electron micrographs. We thank the Hector Fellow Academy, the Helmholtz program Science and Technology of Nanosystems (STN), and the Karlsruhe School of Optics & Photonics (KSOP) for support.

**Author contributions**

J. Q. has fabricated the samples and has performed the measurements and the data processing. M. K. has helped in the simulations and discussions. A. N. has helped in the discussions. M. W. has led the effort and has written the first draft of the paper. All authors have contributed to the final version of the manuscript.

**Additional information**

Reprints and permissions information is available online at www.nature.com/reprints.

Correspondence and requests for materials should be addressed to J. Y. or M. K.

**Competing financial interests**

The authors declare no competing financial interests

**Additional information**

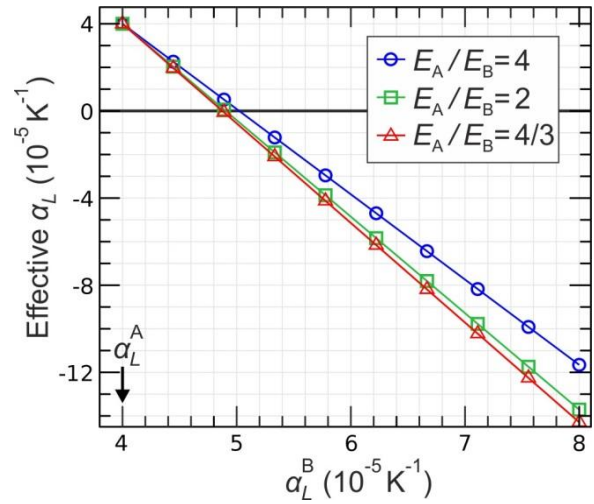

**Supplementary Figure 1 | Effective properties.** Calculated effective metamaterial thermal length-expansion coefficient $\alpha_L$ versus the thermal length-expansion coefficient of constituent material **B**, $\alpha_L^{\text{B}}$. The thermal length-expansion coefficient of constituent material **A** is indicated by the arrow. It has been fixed to $\alpha_L^{\text{A}} = +4 \times 10^{-5}$ K$^{-1}$. The ratio of the Young's moduli $E$ of the two constituent materials is the parameter (see legend). The Young's modulus of constituent material **A** has been fixed to $E_{\text{A}} = 4$ GPa. Note the sign reversal of $\alpha_L$ around $\alpha_L^{\text{B}} \approx +5 \times 10^{-5}$ K$^{-1}$. Geometrical parameters (compare Fig. 1a): $a = 100$ µm, $h = 2.5$ µm, and $L = 40$ µm.